\documentclass[a4paper]{jpconf}
\usepackage{graphicx}
\usepackage{amsmath}
\usepackage{amssymb}
\usepackage[english]{babel}
\begin{document}
\title{Weakly supervised semantic segmentation of tomographic images in the diagnosis of stroke}

\author{A V Dobshik$^1$, A A Tulupov$^{1,2}$ and V B Berikov$^{1,3}$}

\address{$^1$ Novosibirsk State University, Novosibirsk, Russia}
\address{$^2$ International Tomography Center SB RAS, Novosibirsk, Russia}
\address{$^3$ Sobolev Institute of mathematics SB RAS, Novosibirsk, Russia}

\ead{a.dobshik@alumni.nsu.ru, taa@tomo.nsc.ru, berikov@math.nsc.ru}

\begin{abstract}
This paper presents an automatic algorithm for the segmentation of areas affected by an acute stroke on the non-contrast computed tomography brain images. The proposed algorithm is designed for learning in a weakly supervised scenario when some images are labeled accurately, and some images are labeled inaccurately. Wrong labels appear as a result of inaccuracy made by a radiologist in the process of manual annotation of computed tomography images. We propose methods for solving the segmentation problem in the case of inaccurately labeled training data. We use the U-Net neural network architecture with several modifications. Experiments on real computed tomography scans show that the proposed methods increase the segmentation accuracy. 
\end{abstract}

\section{Introduction}
According to the World Health Organization, stroke is the second leading cause of death in the world. It accounts for about 11\% of all deaths. The early diagnosis of acute stroke is of primary importance for deciding on a method for further treatment.

In the process of diagnosing an acute stroke, a radiology specialist performs a manual search and identiﬁcation of the pathological density area in the substance of the brain on the computed tomography (CT) brain images. It is worth noting that non-contrast CT scans are more readily available and have no contraindications, unlike Magnetic Resonance Imaging (MRI) or contrast CT. However, interpretation of non-contrast CT images is rather diﬃcult and ambiguous, as it depends on technical and human factors. An automated system for the recognition of acute stroke on non-contrast CT brain images seems to be a promising solution to this problem. Such a system can be used by radiologists to check the accuracy of their stroke area predictions and to assist in the decision-making process for further treatment.

Many works are devoted to the segmentation of CT images and show a good performance. But in most cases, such algorithms require a large amount of accurately annotated data, which is not easy to obtain in the medical field. For example, the manual segmentation of the stroke areas on CT scans is time-consuming, requires a highly qualified specialist, and therefore is expensive. Moreover, it is known that even a highly skilled specialist can make erroneous predictions with a fairly high probability. If there is a large amount of data, a specialist can label the affected area approximately, for example, with a bounding box or an oval. Such labeling is considered inaccurate because it contains many inaccurately labeled pixels, but it is much easier and faster to get.

In this regard, it is very important to have an algorithm capable of learning from inaccurately labeled data. When constructing a neural network algorithm for weakly supervised learning, it is necessary to take into account that the data on which an algorithm will be trained contain erroneous labels, which will affect the segmentation accuracy on the validation set.

In this work, a neural network algorithm based on U-net \cite{U-Net} architecture is developed and methods for solving the segmentation problem in the case of incomplete information in training data are proposed.

In the rest of the paper, we give a brief overview of the related work, then describe the existing dataset, the neural network architecture used, its modifications, and training details. Then we present the proposed methods for weakly supervised learning, the obtained results, and some implementation details.

\section{Related work}
It is worth noting that in the problems of semantic segmentation of medical images fully convolutional neural networks (CNNs) show a better performance in comparison with classical machine learning methods such as, for example, kNN, SVM, Random Forest, and Adaboost classifiers \cite{Nedelko}. 

Most approaches to solving the weakly supervised segmentation problem can be divided into three groups: 1) a two-stage process where images are initially processed to obtain more accurate segmentation masks and then are fed to the neural network, 2) directly modifying the neural network architecture, 3) a mixture of 1) and 2). An example to the first approach is \cite{renal tumor}. At first, the authors get pseudo-masks from the bounding-box segmentation masks using ConvCRF \cite{ConvCRF} (modification of Fully Connected CRFs \cite{CRF}), then they train the ensemble of CNNs on them. From the ensemble's predictions, a voxel-wise weight map is obtained. These weights are then used in the loss function when training the final CNN. An example to the second approach is \cite{BB-UNet}. In this work, a proposed BBConV layer is added to U-Net, which receives a bounding filter as input. In each skip connection, the intersection between the level contracting layer output and the BBConV layer output is then obtained and further concatenated with the features from the up-sampling layers. Such a technique allows the network to enhance its estimation of where an organ can be. And an example to the third approach is \cite{co-segmentation}. In this work, pseudo-masks are obtained from RECIST diameters using GrabCut \cite{Grabcut}, and then a co-segmentation CNN model with attention modules is trained on them. There are also multi-fidelity methods \cite{multifidelity method}, which are very popular nowadays. Using this approach, it may be assumed that segmentation masks come from the models of different fidelities. Thus, it is much cheaper to generate the sample set, since it may contain only a few high-fidelity and a lot of low-fidelity samples.    

Our approach is inspired by the idea of a weighted loss function in \cite{renal tumor}, but we obtain the weights in another way, and they have a different meaning.

\section{Materials}
To solve our problem, 42 CT brain images of patients with diagnosed acute stroke were obtained from the database of the International Tomography Center SB RAS, Novosibirsk, Russia. In each image, the radiologist performed a careful manual segmentation of areas affected by stroke. These images also contain a certain percentage of inaccurately labeled pixels, but it is relatively small, so we simulate the inaccuracy on the segmentation masks with ovals, see an example in Fig. 1.

\begin{figure}[h]
\begin{minipage}{7.2pc}
\center{\includegraphics[width=8.8pc]{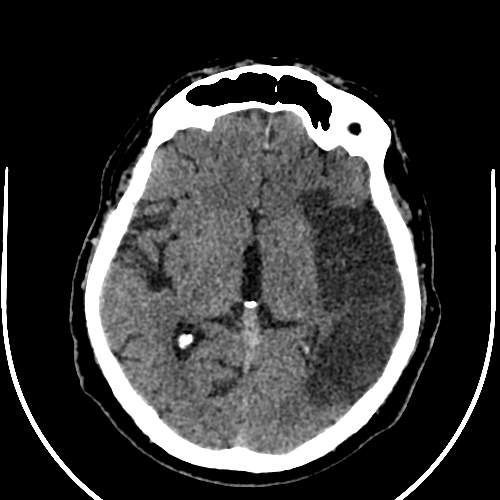} \\ (a)}
\end{minipage}\hspace{2pc}
\begin{minipage}{7.2pc}
\center{\includegraphics[width=8.8pc]{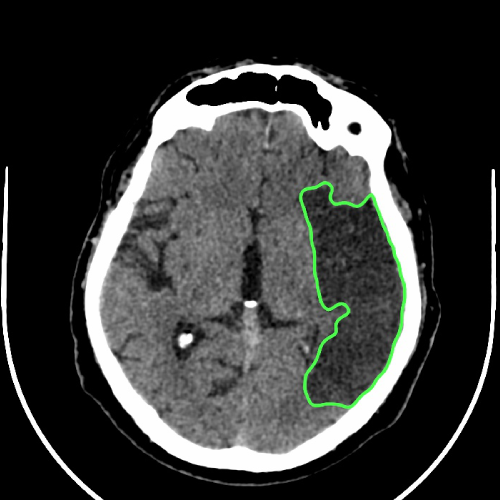} \\ (b)}
\end{minipage}\hspace{2pc}
\begin{minipage}{7.2pc}
\center{\includegraphics[width=8.8pc]{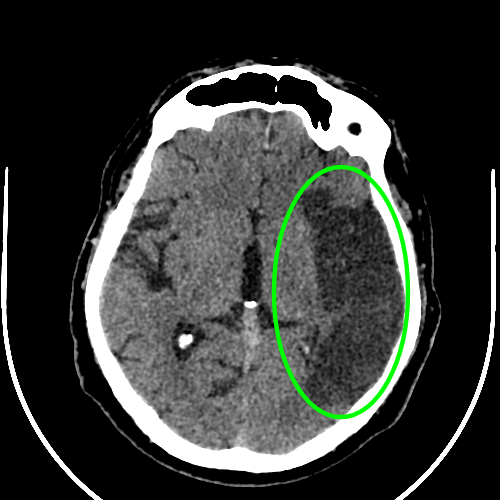} \\ (c)}
\end{minipage} 
\hspace{2pc}%
\begin{minipage}[b]{10pc}\caption{(a) CT brain image, (b) stroke area annotated by a radiologist, (c) simulated inaccurate oval mask.}
\end{minipage}
\end{figure}

We form a new dataset by replacing some $k$ randomly chosen from all 42 segmentation maps made by the radiologist with simulated inaccurate oval masks. In this paper, the case $k=8$ is considered (about 20\% of the dataset). We use such data augmentation techniques as horizontal flip, rotation, random sized cropping, and elastic transformation. All images were prior cropped by the central area and resized to 512 $\times$ 512. Mini-max normalization was applied to CT scans.
 
\section{Neural network architecture}
We use the U-Net architecture with the size of the initial feature channel equal to 32 instead of 64 as in the original work.  We add $padding=1$ to all feature maps before a 3$\times$3 convolutional layer; thus, the network's output is the same size as the input image. Batch normalization and dropout layers are added after each convolutional layer. We add the pyramid pooling module \cite{pp1,pp2} in the bottleneck of U-Net; it helps to capture global information from different regions of the input image. Also, we add another convolutional layer after pyramid pooling, similar to the U-Net convolutional block, see Fig. 2. As is known, such a modification can increase the accuracy \cite{pp1,pp2}. 

\begin{figure}[h!]
\includegraphics[width=6.2in]{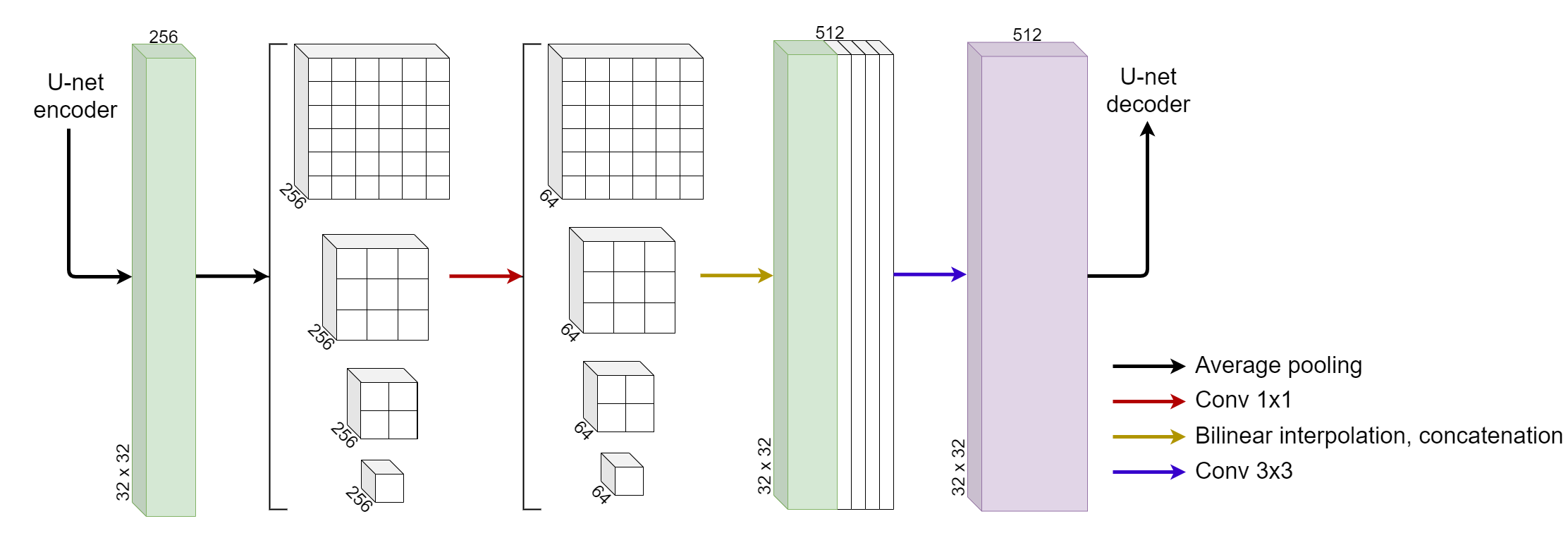}
\caption{\label{Fig. 1} Pyramid pooling module.}
\end{figure}

\section{Training}
We compute the exponential weighted average of the CNN's parameters obtained during learning on the training sample:
\begin{equation}
\begin{gathered}
\tilde{\theta}_0 = \theta_0 \\
\tilde{\theta}_t = \beta\tilde{\theta}_t + (1-\beta)\theta_t,~ ~ t>0 ,~ ~ \beta \in (0,1)
\end{gathered}
\end{equation}
where $t$ is the number of the last mini-batch. Thus, the loss function decreases more smoothly and finds the minimum more accurately. We use the weighted binary cross-entropy loss function as classes are highly imbalanced. The area affected by stroke occupies a small part of the entire image. We also combine it with the Dice loss \cite{loss} by parameter $\alpha \in (0,1)$:
\begin{equation}
L = \alpha{\sum_{i=1}^{N} w_i [-y_i \log p_i - (1 - y_i) \log (1-p_i)] + (1-\alpha)\left(1 - \frac{2\sum_{i}^{N} p_i y_i + \epsilon}{\sum_{i}^{N} p_i^{2} + \sum_{i}^{N} y_i^{2} + \epsilon}\right)}
\end{equation}
\begin{equation}
w_i =  \begin{cases}
{\frac{N_1}{N_0}}, & y_i = 0    \\
1, &  y_i = 1
\end{cases}
\end{equation}
where $N_1$ is the number of all pixels corresponding to all affected areas, $N_0$ is the number of all background pixels, $y_i$ is the ground-truth label of $i$-th pixel, where 1 denotes the stroke area, 0 denotes the background.

\section{Proposed methods for weakly supervised segmentation}
To solve the weakly supervised segmentation problem, we introduce two models of inaccuracy presented in subsections 6.1 and 6.2 respectively. Each model of inaccuracy shows the likelihood that the pixel label is correct. The resulting values obtained from these models are used as weights in the loss function in the manner described in subsection 6.3.

\subsection{The First Model of Inaccuracy}
The method is based on the observation that the farther pixel corresponding to the affected area in the oval mask from the center of the supposed affected area (i.e. the oval) is, the more likely that its label is wrong. Since a specialist makes a coarse-grained annotation, then most of the inaccurately labeled pixels are likely to be near the border of the resulting predicted stroke area.

Suppose $Y=\left\{ y_{ij} \right\}$ is a matrix corresponding to the inaccurate oval segmentation mask, where $i,j =1,\ldots,n$, $n$ is image dimension, $\forall i,j ~ y_{ij} \in \left\{0,1\right\}$, 1 denotes the stroke area, 0 denotes the background. The value obtained from the first model of inaccuracy for a pixel $y_{ij}$ is computed using the Euclidean distance from the center of the supposed stroke area using the following equation:
\begin{equation}
\varphi(y_{ij}) =  \begin{cases}
 1 - \frac{\sqrt{(i-i_0)^2 + (j-j_0)^2}}{\max_{i,j}\sqrt{(i-i_0)^2 + (j-j_0)^2} + \epsilon} , & y_{ij} = 1    \\
1, &  y_{ij} = 0
\end{cases}
\end{equation}
where $i,j$ are matrix indices of pixel $y_{ij}$, and $i_0,j_0$ are matrix indices of the central pixel of the oval.~$\epsilon$ is a small positive real number; thus, such a normalization by $\max_{i,j}\sqrt{(i-i_0)^2 + (j-j_0)^2} + \epsilon$ makes $\varphi(y_{ij}) \in (0,1]$ for all $y_{ij}=1$. 

\subsection{The Second Model of Inaccuracy}
This model is also based on the observation that the farther pixel corresponding to the affected area in the oval mask from the center of the supposed affected area is, the more likely that its label is wrong. The difference of the second model is that we use the Mahalanobis distance. Mahalanobis distance differs from Euclidean distance in that it takes into account the correlation between variables and is scale invariant. Mahalanobis distance is computed by the following equation:
\begin{equation} 
D^2(y_{ij}) =  (y-\mu)^{T}C^{-1}(y-\mu)
\end{equation} 
where $y=(i,j)$ are the indices of pixel $y_{ij}$, $\mu$ is a vector of mean values of $i$ and $j$ indices of pixels corresponding to the affected area. $C^{-1}$ is the inverse covariance matrix calculated from indices of pixels belonging to the oval.
The values obtained from the second model of inaccuracy are calculated similarly:
\begin{equation}
\varphi(y_{ij}) =  \begin{cases}
 1 - \frac{D(y_{ij})}{\max_{i,j} D(y_{ij}) + \epsilon} , & y_{ij} = 1    \\
1, &  y_{ij} = 0
\end{cases}
\end{equation}

The distribution of the values obtained from the first and the second models of inaccuracy is shown in Fig. 3. 
\begin{figure}[h]
\begin{minipage}{7.2pc}
\center{\includegraphics[width=8.8pc]{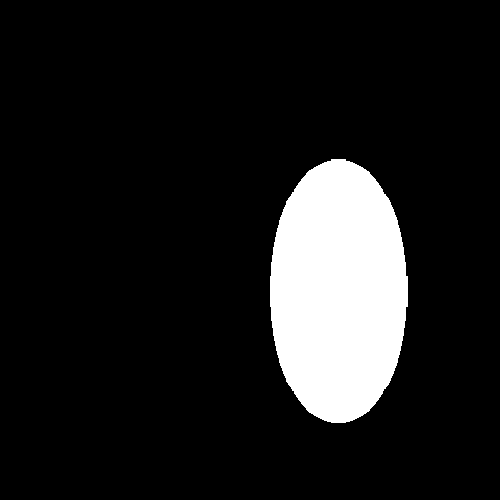} \\ (a)}
\end{minipage}\hspace{2pc}
\begin{minipage}{7.2pc}
\center{\includegraphics[width=8.8pc]{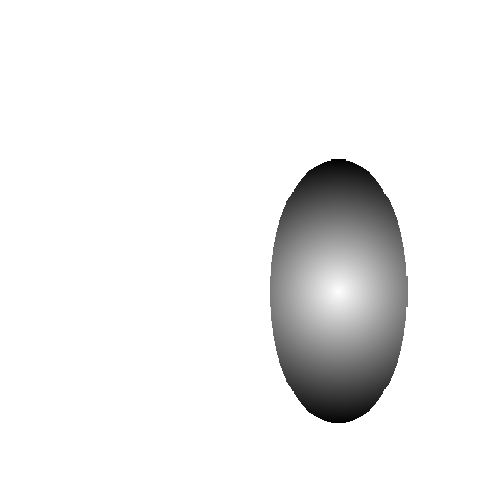} \\ (b)}
\end{minipage}\hspace{2pc}
\begin{minipage}{7.2pc}
\center{\includegraphics[width=8.8pc]{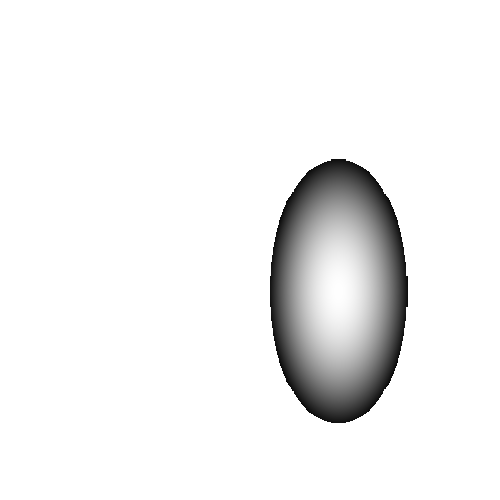} \\ (c)}
\end{minipage} 
\hspace{2pc}%
\begin{minipage}[b]{10pc}\caption{(a) binary oval mask, (b) the 1st Model of Inaccuracy, (c) the 2nd Model of Inaccuracy.}
\end{minipage}
\end{figure}
The darker the pixel, the lower its value of $\varphi(y_{ij})$ is.

\subsection{Modified Loss Function}
The value for the $i$-th pixel obtained by the first or the second model of inaccuracy is then used in loss function as weight $\varphi_i$:
\begin{equation}
\tilde{L} = {\sum_{i=1}^{N} w_i [-\varphi_i y_i \log p_i - (1 - y_i) \log (1-p_i)]}
\end{equation}

White pixels are less reliable in oval masks. During the training stage, multiplying by weights has less impact on the loss function contributing. This also results in less impact on parameter changes by the backpropagation method. It is worth noting that the weights $\varphi_i$ in the loss function are only used for $k$ inaccurate oval segmentation masks.

\section{Results of experiments}
To make the results more reliable, we run the experiments on five different subsets of data and average the results obtained by training on them. We form these subsets by replacing $k$ different segmentation maps made by the radiologist with simulated inaccurate oval masks without repetitions. Dice similarity coefficient (DSC) and 5-fold cross-validation testing were used.

In order to reduce the impact of inaccurately labeled pixels, we convert the values $\varphi_i$ by the power function $f(x)=x^n$. The results of applying the First and Second Models of Inaccuracy (MoI) are shown in Table 1.  It shows the DSC value and its standard deviation depending on the power $n$ in the function $f$.
\begin{table}[h]
\caption{\label{ex}Comparison of the mean values and standard deviations for 1st and 2nd MoI.}
\begin{center}
\begin{tabular}{lll}
\br
$n$ & 1st MoI & 2nd MoI \\
\mr
1 & \textbf{0.7609} $\pm0.012$ & 0.7523 $\pm0.014$\\
1.5 & 0.7573 $\pm0.012$ &  0.7575 $\pm0.012$ \\
2 & 0.7534 $\pm0.013$ & \textbf{0.7633} $\pm0.014$ \\
\br
\end{tabular}
\end{center}
\end{table}

The result of the experiment when training without weights $\varphi_i$ in the loss function is \textbf{0.7478}$\pm0.015$ DSC. Thus, the proposed methods with the first and the second models of inaccuracy improve the segmentation quality on average by \textbf{1.31}$\%$ and \textbf{1.55}$\%$, respectively.

\section{Implementation details}
All models are implemented in PyTorch. Training is
conducted with a batch size of 4 images for a total of 280 epochs. We use the Adam \cite{Adam} optimizer and the one cycle learning rate training policy \cite{oclr} with the maximum learning rate of 0.001 and a weight decay of
0.0005 for L2 regularization. We used the parameter of exponential smoothing $\beta=0.995$, the parameter in loss function $\alpha=0.5$ and dropout layer probability $p=0.4$. In the 1st and 2nd MoI $\epsilon=1$.

\section{Conclusion}
In this work, the problem of weakly supervised semantic segmentation of non-contrast computed tomographic brain images in the diagnosis of stroke was considered. Under the weakly supervised task, we understand the scenario, when some images are labeled accurately and some images are labeled inaccurately. This task is important since accurately annotated data is expensive and not easy to obtain. 

Our proposed methods for weakly supervised segmentation using weights obtained by first and second models of inaccuracy improve the quality of segmentation; their effectiveness has been tested on real computed tomography images. In the future, it is planned to use other different neural network architectures, for example, 3D U-Net \cite{3D U-Net}.

\ack{The work was partly supported by RFBR grant 19-29-01175.}

\smallskip

\end{document}